\documentclass[a4paper]{jpconf}
\usepackage{graphicx}

\usepackage{hyperref}
\usepackage{fancyhdr}
\fancypagestyle{firstpage}{%
  \lhead{\footnotesize{ V. Voronenkov \textit{et al} 2019 \textit{J. Phys.: Conf. Ser.} \textbf{1199} 012004  }}
  	
  \cfoot{}
	\lfoot{}
  \rhead{\footnotesize{\href{https://doi.org/10.1088/1742-6596/1199/1/012004}{doi:10.1088/1742-6596/1199/1/012004}}}
}

\usepackage{cite}
\begin{document}

\title{Free-standing 2-inch bulk GaN crystal fabrication by HVPE using a carbon buffer layer}

\author{
Vladislav~Voronenkov$^{1,2}$,
Andrey~Leonidov$^{3}$,
Natalia~Bochkareva$^{1}$, 
Ruslan~Gorbunov$^{1,2}$,
Philippe~Latyshev$^{4}$,
Yuri~Lelikov$^{1,2}$,
Viktor~Kogotkov$^{2}$,
Andrey~Zubrilov$^{1,2}$ and
Yuri~Shreter$^{1,2}$}

\address{$^{1}$Ioffe Institute,  Politehnicheskaya~26, St.~Petersburg,  194021, Russia}
\address{$^{2}$TRINITRI-Technology LLC, 194223, St. Petersburg, Russia}
\address{$^{3}$Peter the Great St.~Petersburg Polytechnic University,  Politehnicheskaya~29, St.~Petersburg, 195251, Russia}
\address{$^{4}$OOO NTS, 197198, St. Petersburg, Russia}

\ead{voronenkov@mail.ioffe.ru}

\begin{abstract}
A free-standing bulk gallium nitride layer with a thickness of 365~$\mu$m and a diameter of 50~mm was obtained by hydride vapor phase epitaxy on a sapphire substrate with a carbon buffer layer. The carbon buffer layer was deposited by thermal decomposition of methane \textit{in~situ} in the same process with the growth of a bulk GaN layer. The bulk GaN layer grown on the carbon buffer layer self-separated from the sapphire substrate during the cooling after the growth. The dislocation density  was $8\cdot10^{6}$~cm$^{-2}$. The (0002) X-Ray rocking curve full width at half maximum  was 164 arcsec.
\end{abstract}
\thispagestyle{firstpage}

\section{Introduction}
Today,  HVPE heteroepitaxial growth of GaN on foreign substrates is the  main method for commercial production of GaN substrates. An important step in this technology is the separation of the bulk GaN layer from the substrate after the growth process. Thick GaN layers with thicknesses of several hundred microns typically separate from sapphire substrates during the cooling  after the growth, but this self-separation process is accompanied by cracking of the GaN layer \cite{Natural-Stress-Moustakas2007,YAMANE20121,ASHRAF20102-FS-GaN-selfseparation}. The cracking of a GaN layer can be suppressed by growing a bulk GaN layer with a thickness of several millimeters, or by reducing the bonding energy between a substrate and a GaN layer.
Different methods were proposed to reduce the bonding energy between a substrate and a GaN layer, including porous layers created by dry etching  \cite{H2-porous-etching-YEH2011}, wet etching \cite{LEE-KOH-POROUS-ETCH},  electrochemical etching~\cite{Mynbaeva-porous-liftoff} or GaN decomposition induced by a TiN mask \cite{USUI-VAS-2003}; epitaxial lateral overgrowth over dielectric masks~\cite{gogova2004elog,lipski2010fabrication, HENNIG2008911-WSiN-ELOG, AMILUSIK201399}, employing substrates with a cleavage plane parallel to the c-axis of GaN \cite{Ga2O3-lift-off-Gogova2012, SCAM-Matsuoka-2017}, and weakly bonded buffer  layers  \cite{USUI-TiC-2012, Graphene-IBM-kim2014principle}.
All these methods either require using non-standard substrates  \cite{Ga2O3-lift-off-Gogova2012, SCAM-Matsuoka-2017} or special \textit{ex~situ} pre-growth processing of the substrate like etching, buffer layer deposition, dielectric or metal mask fabrication, or GaN template structure growth.

\begin{figure}
\begin{center}
\includegraphics[width=\textwidth]{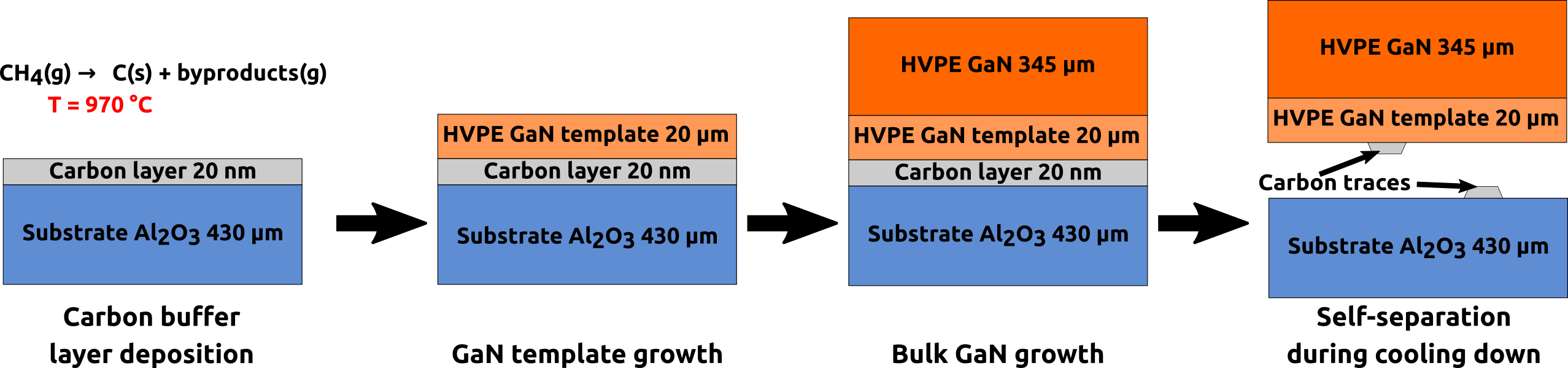}
\end{center}
\caption{\label{SCHEME}Schematic representation of the free-standing GaN  fabrication on a sapphire substrate with a carbon buffer layer.}
\end{figure}

Carbon in the form of graphite  is a promising material for creating buffer layers that facilitate self-separation due to the hexagonal symmetry of the crystal lattice and a low interlayer binding energy \cite{Liu2012-graphite-binding}.
Earlier, a nanocrystalline graphite carbon buffer layer deposited by PECVD on a sapphire substrate  was  used for GaN growth and self-separation. A 200-$\mu$m thick GaN layer was grown and free-standing GaN pieces were obtained \cite{CARBON-Phil-2016}.
In this work, a bulk GaN layer with a thickness of~365~$\mu$m  was grown on a sapphire substrate with a carbon buffer layer deposited \textit{in~situ}, and a wafer-scale  self-separation has been demonstrated (figure \ref{SCHEME}).
\section{Experimental details}

The experiment used a standard epi-polished 2-inch sapphire substrate with a thickness of 430~$\mu$m and a crystallographic orientation of (0001) with a miscut of 0.5$^{\circ}$ towards the m-plane.  

A carbon buffer layer and a GaN film with a thickness of 20~$\mu$m were grown in a multi-wafer HVPE reactor with an installed methane supply line. The design of the reactor limited the maximal GaN layer thickness to 200~$\mu$m; therefore, to grow the bulk layer, the structure was transferred to the vertical HVPE reactor optimized for the bulk layer deposition.

A carbon layer was deposited using the methane thermal decomposition process from a CH$_4$/H$_2$ mixture~\cite{BECKER1998-ch4-h2} at a deposition temperature of  1020~$^{\circ}$C, a total process pressure of 105~kPa and a methane partial pressure of 1.3~kPa. The  deposition rate was 10~nm/min.
After that, a 20-$\mu$m thick GaN layer was deposited at a temperature of 960~$^{\circ}$C. Then,  the structure was cooled down and transferred to a vertical HVPE reactor  where a 345-$\mu$m thick GaN layer was grown at a growth rate of~100~$\mu$m/h, a V/III~ratio of 35, a total reactor pressure of 15~kPa, and a substrate temperature ramping from 880~$^{\circ}$C at the beginning of the growth process to 920~$^{\circ}$C at the end of the growth process. Self-separation induced by thermal stress occurred during the cooling down after the growth process, and a single piece GaN layer with a thickness of 365~$\mu$m and a diameter of  50~mm was obtained.

\section{Results and discussion} 
\begin{figure}
\begin{center}
\includegraphics[width=\textwidth]{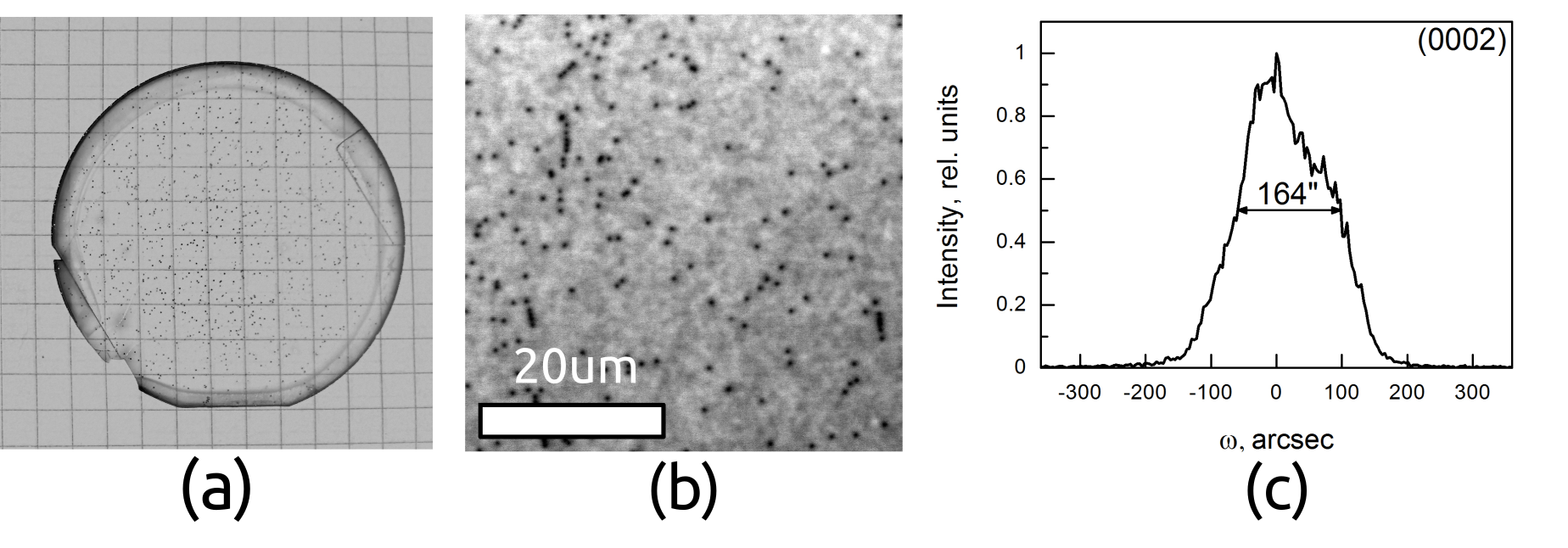}
\end{center}
\caption{\label{BULK-XRAY-CL} 
(a) A photograph of  free-standing GaN layer with a thickness of 365~$\mu$m and a diameter of 50~mm. The surface is optically smooth and the V-shaped pit density is below 40~cm$^{-2}$.
(b) A cathodoluminescence micrograph of the Ga-face surface. The dark spot density is $8\cdot10^{6}$ cm$^{-2}$.
(c)  A (0002) XRD rocking curve of the Ga face of free-standing GaN. 
} 
\end{figure}

A photograph of a free-standing GaN layer is shown in figure~\ref{BULK-XRAY-CL}(a). The surface of the layer is smooth, the V-shaped pit density is 40~cm$^{-2}$, and the opening angle of V-shaped pits is in the range of 80$^{\circ}$-100$^{\circ}$ that is typical for  GaN layers grown in the low-temperature mode \cite{VVVvoronenkov2013two}. 

The (0002) XRD rocking curve of the Ga face of the free-standing GaN is shown in figure~\ref{BULK-XRAY-CL}(c). The FWHM value is 164~arcsec. 
The lattice curvature radius of free-standing GaN measured by XRD is 3.1~m. The residual bow is typical for free-standing GaN layers obtained by heteroepitaxy and is a result of a built-in strain caused by  defect density gradients along the c-axis of a GaN layer \cite{Lucnik2011residual-bow,LIPSKI2012-wafer-bow} and by the inclination of threading dislocations \cite{foronda-speck2016inclined-dislocation-bow}.

The threading dislocation density on the Ga-face surface of the free-standing  GaN layer was estimated to be $8\cdot10^{6}$~cm$^{-2}$ (figure~\ref{BULK-XRAY-CL}(b)),  which is typical for GaN layers with thicknesses of 350-400~$\mu$m, grown on a sapphire substrate \cite{FUJITO2009-5mm}. The estimation was made by measuring the dark spot density using the cathodoluminescent microscopy~\cite{Sugahara1999-cl-tem-disloactions}.

\begin{figure}
\begin{center}
\includegraphics[width=\textwidth]{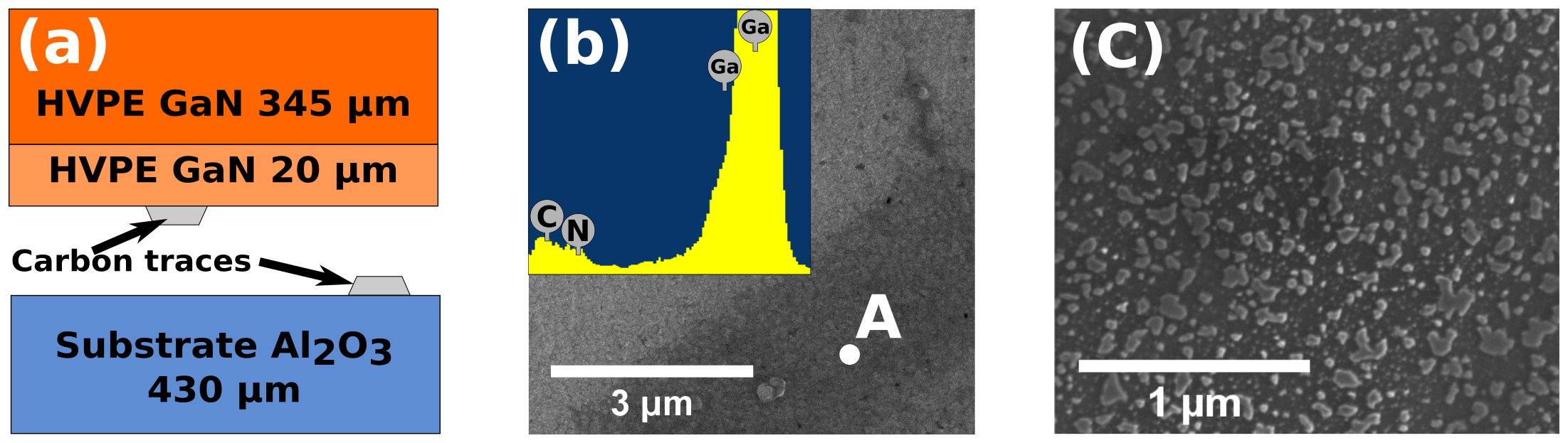}
\end{center}
\caption{\label{SEM-EDX} 
(a) Schematic representation of a bulk GaN layer  and a growth substrate after self-separation. 
(b) A SEM micrograph  of the N-face of a GaN layer. The inset shows the EDX spectrum of the surface, measured at point A and demonstrating remnants of the carbon buffer layer. 
(c) A SEM micrograph of the sapphire substrate surface shows submicron-sized irregularities formed during the process of free-standing GaN fabrication.}
\end{figure}

The surfaces of the N-face of the GaN layer and the sapphire substrate were smooth and mirror-like. Both these surfaces were examined by scanning electron microscopy combined with energy-dispersive X-ray spectroscopy. The carbon buffer layer remnants were found on the N-face of free-standing GaN layer (figure~\ref{SEM-EDX}(b)) and the sapphire substrate surfaces (not shown). This confirms that the self-separation occurred strictly along the carbon buffer layer. The sapphire substrate surface was covered by submicron-sized irregularities (figure~\ref{SEM-EDX}(c)), while no irregularities of the same scale were observed on the GaN surface. A possible origin of  such irregularities could be the reduction of sapphire by carbon \cite{FOSTER-Al2O3-C}. Sapphire etching and the resulting cavity formation may also lower the bonding energy between a GaN layer and a substrate and facilitate the self-separation process.

\section{Conclusion}
A bulk GaN layer was epitaxially grown on a sapphire substrate with a carbon buffer layer. The layer spontaneously separated from the substrate along the carbon buffer layer during the cooling down process, and a free-standing GaN layer with a diameter of 50 mm and a thickness of 365~$\mu$m was obtained.
The surface morphology and crystalline structure of the free-standing GaN layer did not degrade compared to a GaN layer grown on a bare sapphire substrate in similar conditions.

\section{Acknowledgments}
The authors gratefully thank O.~Mededev and O.~Vyvenko from St.~Petersburg State University for help with microscopic investigations.

\section*{References}
\bibliography{self-separation}

\begin{thebibliography}{10}
\providecommand{\url}[1]{#1}
\csname url@samestyle\endcsname
\providecommand{\newblock}{\relax}
\providecommand{\bibinfo}[2]{#2}
\providecommand{\BIBentrySTDinterwordspacing}{\spaceskip=0pt\relax}
\providecommand{\BIBentryALTinterwordstretchfactor}{4}
\providecommand{\BIBentryALTinterwordspacing}{\spaceskip=\fontdimen2\font plus
\BIBentryALTinterwordstretchfactor\fontdimen3\font minus
  \fontdimen4\font\relax}
\providecommand{\BIBforeignlanguage}[2]{{%
\expandafter\ifx\csname l@#1\endcsname\relax
\typeout{** WARNING: IEEEtran.bst: No hyphenation pattern has been}%
\typeout{** loaded for the language `#1'. Using the pattern for}%
\typeout{** the default language instead.}%
\else
\language=\csname l@#1\endcsname
\fi
#2}}
\providecommand{\BIBdecl}{\relax}
\BIBdecl

\bibitem{Natural-Stress-Moustakas2007}
\BIBentryALTinterwordspacing
A.~D. Williams and T.~Moustakas, ``{Formation of large-area freestanding
  gallium nitride substrates by natural stress-induced separation of GaN and
  sapphire},'' \emph{Journal of Crystal Growth}, vol. 300, no.~1, pp. 37 -- 41,
  2007, first International Symposium on Growth of Nitrides. [Online].
  Available: \url{https://doi.org/10.1016/j.jcrysgro.2006.10.224}
\BIBentrySTDinterwordspacing

\bibitem{YAMANE20121}
\BIBentryALTinterwordspacing
K.~Yamane, M.~Ueno, H.~Furuya, N.~Okada, and K.~Tadatomo, ``{Successful natural
  stress-induced separation of hydride vapor phase epitaxy-grown GaN layers on
  sapphire substrates},'' \emph{Journal of Crystal Growth}, vol. 358, pp. 1 --
  4, 2012. [Online]. Available:
  \url{https://doi.org/10.1016/j.jcrysgro.2012.07.038}
\BIBentrySTDinterwordspacing

\bibitem{ASHRAF20102-FS-GaN-selfseparation}
\BIBentryALTinterwordspacing
H.~Ashraf, R.~Kudrawiec, J.~Weyher, J.~Serafinczuk, J.~Misiewicz, and
  P.~Hageman, ``{Properties and preparation of high quality, free-standing GaN
  substrates and study of spontaneous separation mechanism},'' \emph{Journal of
  Crystal Growth}, vol. 312, no.~16, pp. 2398 -- 2403, 2010. [Online].
  Available: \url{https://doi.org/10.1016/j.jcrysgro.2010.05.004}
\BIBentrySTDinterwordspacing

\bibitem{H2-porous-etching-YEH2011}
\BIBentryALTinterwordspacing
Y.-H. Yeh, K.-M. Chen, Y.-H. Wu, Y.-C. Hsu, T.-Y. Yu, and W.-I. Lee,
  ``{Hydrogen etching of GaN and its application to produce free-standing GaN
  thick films},'' \emph{Journal of Crystal Growth}, vol. 333, no.~1, pp. 16 --
  19, 2011. [Online]. Available:
  \url{https://doi.org/10.1016/j.jcrysgro.2011.08.022}
\BIBentrySTDinterwordspacing

\bibitem{LEE-KOH-POROUS-ETCH}
\BIBentryALTinterwordspacing
C.-Y. Lee, Y.-P. Lan, P.-M. Tu, S.-C. Hsu, C.-C. Lin, H.-C. Kuo, G.-C. Chi, and
  C.-Y. Chang, ``{Natural substrate lift-off technique for vertical
  light-emitting diodes},'' \emph{Applied Physics Express}, vol.~7, no.~4, p.
  042103, 2014. [Online]. Available:
  \url{https://doi.org/10.7567/APEX.7.042103}
\BIBentrySTDinterwordspacing

\bibitem{Mynbaeva-porous-liftoff}
\BIBentryALTinterwordspacing
M.~G. Mynbaeva, A.~E. Nikolaev, A.~A. Sitnikova, and K.~D. Mynbaev, ``{HVPE
  homo-epitaxial growth of GaN on porous substrates},'' \emph{CrystEngComm},
  vol.~15, pp. 3640--3646, 2013. [Online]. Available:
  \url{http://dx.doi.org/10.1039/C3CE27099H}
\BIBentrySTDinterwordspacing

\bibitem{USUI-VAS-2003}
\BIBentryALTinterwordspacing
Y.~Oshima, T.~Eri, M.~Shibata, H.~Sunakawa, K.~Kobayashi, T.~Ichihashi, and
  A.~Usui, ``{Preparation of Freestanding GaN Wafers by Hydride Vapor Phase
  Epitaxy with Void-Assisted Separation},'' \emph{Japanese Journal of Applied
  Physics}, vol.~42, no.~1A, p.~L1, 2003. [Online]. Available:
  \url{https://doi.org/10.1143/JJAP.42.L1}
\BIBentrySTDinterwordspacing

\bibitem{gogova2004elog}
\BIBentryALTinterwordspacing
D.~Gogova, A.~Kasic, H.~Larsson, C.~Hemmingsson, B.~Monemar, F.~Tuomisto,
  K.~Saarinen, L.~Dobos, B.~Pécz, P.~Gibart, and B.~Beaumont, ``{Strain-free
  bulk-like GaN grown by hydride-vapor-phase-epitaxy on two-step epitaxial
  lateral overgrown GaN template},'' \emph{Journal of Applied Physics},
  vol.~96, no.~1, pp. 799--806, 2004. [Online]. Available:
  \url{https://doi.org/10.1063/1.1753073}
\BIBentrySTDinterwordspacing

\bibitem{lipski2010fabrication}
\BIBentryALTinterwordspacing
F.~Lipski, T.~Wunderer, S.~Schwaiger, and F.~Scholz, ``{Fabrication of
  freestanding 2"-GaN wafers by hydride vapour phase epitaxy and
  self-separation during cooldown},'' \emph{Physica Status Solidi (a)}, vol.
  207, no.~6, pp. 1287--1291, 2010. [Online]. Available:
  \url{https://doi.org/10.1002/pssa.200983517}
\BIBentrySTDinterwordspacing

\bibitem{HENNIG2008911-WSiN-ELOG}
\BIBentryALTinterwordspacing
C.~Hennig, E.~Richter, M.~Weyers, and G.~Trankle, ``{Freestanding 2-in GaN
  layers using lateral overgrowth with HVPE},'' \emph{Journal of Crystal
  Growth}, vol. 310, no.~5, pp. 911 -- 915, 2008, proceedings of the E-MRS
  Conference, Symposium G. [Online]. Available:
  \url{https://doi.org/10.1016/j.jcrysgro.2007.11.102}
\BIBentrySTDinterwordspacing

\bibitem{AMILUSIK201399}
\BIBentryALTinterwordspacing
M.~Amilusik, T.~Sochacki, B.~Lucznik, M.~Bockowski, B.~Sadovyi, A.~Presz,
  I.~Dziecielewski, and I.~Grzegory, ``{Analysis of self-lift-off process
  during HVPE growth of GaN on MOCVD-GaN/sapphire substrates with
  photolitographically patterned Ti mask},'' \emph{Journal of Crystal Growth},
  vol. 380, pp. 99 -- 105, 2013. [Online]. Available:
  \url{https://doi.org/10.1016/j.jcrysgro.2013.06.005}
\BIBentrySTDinterwordspacing

\bibitem{Ga2O3-lift-off-Gogova2012}
\BIBentryALTinterwordspacing
K.~Kachel, M.~Korytov, D.~Gogova, Z.~Galazka, M.~Albrecht, R.~Zwierz, D.~Siche,
  S.~Golka, A.~Kwasniewski, M.~Schmidbauer, and R.~Fornari, ``{A new approach
  to free-standing GaN using $\beta$-Ga$_2$O$_3$ as a substrate},''
  \emph{CrystEngComm}, vol.~14, pp. 8536--8540, 2012. [Online]. Available:
  \url{http://dx.doi.org/10.1039/C2CE25976A}
\BIBentrySTDinterwordspacing

\bibitem{SCAM-Matsuoka-2017}
\BIBentryALTinterwordspacing
K.~Ohnishi, M.~Kanoh, T.~Tanikawa, S.~Kuboya, T.~Mukai, and T.~Matsuoka,
  ``{Halide vapor phase epitaxy of thick GaN films on ScAlMgO$_4$ substrates
  and their self-separation for fabricating freestanding wafers},''
  \emph{Applied Physics Express}, vol.~10, no.~10, p. 101001, 2017. [Online].
  Available: \url{https://doi.org/10.7567/APEX.10.101001}
\BIBentrySTDinterwordspacing

\bibitem{USUI-TiC-2012}
\BIBentryALTinterwordspacing
H.~Geng, H.~Sunakawa, N.~Sumi, K.~Yamamoto, A.~A. Yamaguchi, and A.~Usui,
  ``{Growth and strain characterization of high quality GaN crystal by HVPE},''
  \emph{Journal of Crystal Growth}, vol. 350, no.~1, pp. 44 -- 49, 2012, the
  7th International Workshop on Bulk Nitride Semiconductors. [Online].
  Available: \url{https://doi.org/10.1016/j.jcrysgro.2011.12.020}
\BIBentrySTDinterwordspacing

\bibitem{Graphene-IBM-kim2014principle}
\BIBentryALTinterwordspacing
J.~Kim, C.~Bayram, H.~Park, C.-W. Cheng, C.~Dimitrakopoulos, J.~A. Ott, K.~B.
  Reuter, S.~W. Bedell, and D.~K. Sadana, ``{Principle of direct van der Waals
  epitaxy of single-crystalline films on epitaxial graphene},'' \emph{Nature
  communications}, vol.~5, p. 4836, 2014. [Online]. Available:
  \url{https://doi.org/10.1038/ncomms5836}
\BIBentrySTDinterwordspacing

\bibitem{Liu2012-graphite-binding}
\BIBentryALTinterwordspacing
Z.~Liu, J.~Z. Liu, Y.~Cheng, Z.~Li, L.~Wang, and Q.~Zheng, ``Interlayer binding
  energy of graphite: A mesoscopic determination from deformation,''
  \emph{Phys. Rev. B}, vol.~85, p. 205418, May 2012. [Online]. Available:
  \url{https://doi.org/10.1103/PhysRevB.85.205418}
\BIBentrySTDinterwordspacing

\bibitem{CARBON-Phil-2016}
\BIBentryALTinterwordspacing
A.~S. Altakhov, R.~I. Gorbunov, L.~A. Kasharina, F.~E. Latyshev, V.~A. Tarala,
  and Y.~G. Shreter, ``{Amorphous carbon buffer layers for separating free
  gallium nitride films},'' \emph{Technical Physics Letters}, vol.~42, no.~11,
  pp. 1076--1078, Nov 2016. [Online]. Available:
  \url{https://doi.org/10.1134/S106378501611002X}
\BIBentrySTDinterwordspacing

\bibitem{BECKER1998-ch4-h2}
\BIBentryALTinterwordspacing
A.~Becker and K.~Huttinger, ``{Chemistry and kinetics of chemical vapor
  deposition of pyrocarbon - IV pyrocarbon deposition from methane in the low
  temperature regime},'' \emph{Carbon}, vol.~36, no.~3, pp. 213 -- 224, 1998.
  [Online]. Available: \url{https://doi.org/10.1016/S0008-6223(97)00177-2}
\BIBentrySTDinterwordspacing

\bibitem{VVVvoronenkov2013two}
\BIBentryALTinterwordspacing
V.~Voronenkov, N.~Bochkareva, R.~Gorbunov, P.~Latyshev, Y.~Lelikov, Y.~Rebane,
  A.~Tsyuk, A.~Zubrilov, U.~Popp, M.~Strafela, and Y.~Shreter, ``{Two modes of
  HVPE growth of GaN and related macrodefects},'' \emph{Physica Status Solidi
  (c)}, vol.~10, no.~3, pp. 468--471, 2013. [Online]. Available:
  \url{https://doi.org/10.1002/pssc.201200701}
\BIBentrySTDinterwordspacing

\bibitem{Lucnik2011residual-bow}
\BIBentryALTinterwordspacing
B.~Lucznik, T.~Sochacki, M.~Sarzynski, M.~Krysko, I.~Dziecielewski,
  I.~Grzegory, and S.~Porowski, ``{C-plane bowing in free standing GaN crystals
  grown by HVPE on GaN-sapphire substrates with photolithographically patterned
  Ti masks},'' \emph{physica status solidi c}, vol.~8, no. 7‐8, pp.
  2117--2119, 2011. [Online]. Available:
  \url{https://doi.org/10.1002/pssc.201001000}
\BIBentrySTDinterwordspacing

\bibitem{LIPSKI2012-wafer-bow}
\BIBentryALTinterwordspacing
F.~Lipski, M.~Klein, X.~Yao, and F.~Scholz, ``{Studies about wafer bow of
  freestanding GaN substrates grown by hydride vapor phase epitaxy},''
  \emph{Journal of Crystal Growth}, vol. 352, no.~1, pp. 235 -- 238, 2012, the
  Proceedings of the 18th American Conference on Crystal Growth and Epitaxy.
  [Online]. Available: \url{https://doi.org/10.1016/j.jcrysgro.2011.10.021}
\BIBentrySTDinterwordspacing

\bibitem{foronda-speck2016inclined-dislocation-bow}
\BIBentryALTinterwordspacing
H.~M. Foronda, A.~E. Romanov, E.~C. Young, C.~A. Robertson, G.~E. Beltz, and
  J.~S. Speck, ``{Curvature and bow of bulk GaN substrates},'' \emph{Journal of
  Applied Physics}, vol. 120, no.~3, p. 035104, 2016. [Online]. Available:
  \url{https://doi.org/10.1063/1.4959073}
\BIBentrySTDinterwordspacing

\bibitem{FUJITO2009-5mm}
\BIBentryALTinterwordspacing
K.~Fujito, S.~Kubo, H.~Nagaoka, T.~Mochizuki, H.~Namita, and S.~Nagao, ``{Bulk
  GaN crystals grown by HVPE},'' \emph{Journal of Crystal Growth}, vol. 311,
  no.~10, pp. 3011 -- 3014, 2009, proceedings of the 2nd International
  Symposium on Growth of III Nitrides. [Online]. Available:
  \url{https://doi.org/10.1016/j.jcrysgro.2009.01.046}
\BIBentrySTDinterwordspacing

\bibitem{Sugahara1999-cl-tem-disloactions}
\BIBentryALTinterwordspacing
T.~Sugahara, H.~Sato, M.~Hao, Y.~Naoi, S.~Kurai, S.~Tottori, K.~Yamashita,
  K.~Nishino, L.~T. Romano, and S.~Sakai, ``{Direct Evidence that Dislocations
  are Non-Radiative Recombination Centers in GaN},'' \emph{Japanese Journal of
  Applied Physics}, vol.~37, no.~4A, p. L398, 1998. [Online]. Available:
  \url{https://doi.org/10.1143/JJAP.37.L398}
\BIBentrySTDinterwordspacing

\bibitem{FOSTER-Al2O3-C}
\BIBentryALTinterwordspacing
L.~M. Foster, G.~Long, and M.~S. Hutner, ``{Reactions Between Aluminum Oxide
  and Carbon The Al$_2$O$_3$-Al$_4$C$_3$ Phase Diagram},'' \emph{Journal of the
  American Ceramic Society}, vol.~39, no.~1, pp. 1--11. [Online]. Available:
  \url{https://doi.org/10.1111/j.1151-2916.1956.tb15588.x}
\BIBentrySTDinterwordspacing

\end{thebibliography}

\end{document}